\documentclass{article}

\usepackage{PRIMEarxiv}

\usepackage[utf8]{inputenc} 
\usepackage[T1]{fontenc}    
\usepackage{hyperref}       
\usepackage{url}            
\usepackage{booktabs}       
\usepackage{amsfonts}       
\usepackage{nicefrac}       
\usepackage{microtype}      
\usepackage{lipsum}
\usepackage{fancyhdr}       
\usepackage{graphicx}       
\graphicspath{{media/}}     

\title{Automatic Speech Recognition for Biomedical Data in Bengali Language
}

\author{
  Shariar Kabir\thanks{Corresponding author.} \\
  MedAI Limited\\
  \texttt{shariar.kabir@medaihealth.com} \\
     \And
  Nazmun Nahar \\
  MedAI Limited\\
  \texttt{ocean.rahan@medaihealth.com} \\
    \AND
  Shyamasree Saha \\
  MedAI Limited\\
  \texttt{shyama.saha@medaihealth.com} \
    \AND
  Mamunur Rashid\thanks{Corresponding author.} \\
  University of Birmingham\\
  \texttt{m.rashid.1@bham.ac.uk} \\
}

\begin{document}
\maketitle

\begin{abstract}
This paper presents the development of a prototype Automatic Speech Recognition (ASR) system specifically designed for Bengali biomedical data.  Recent advancements in Bengali ASR are encouraging, but a lack of domain-specific data limits the creation of practical healthcare ASR models. This project bridges this gap by developing an ASR system tailored for Bengali medical terms like symptoms, severity levels, and diseases, encompassing two major dialects: Bengali and Sylheti. We train and evaluate two popular ASR frameworks on a comprehensive 46-hour Bengali medical corpus.  Our core objective is to create deployable health-domain ASR systems for digital health applications, ultimately increasing accessibility for non-technical users in the healthcare sector.
\end{abstract}


\section{Introduction}
Recent advancements in domain specific Automated Speech Recognition (ASR) and Large Language Models (LLM), have significantly boosted the adoption of AI in digital services across many different industries such as financial service, healthcare. In the healthcare industry in particular, integration of AI-driven solutions such as conversational chatbots, voice interactive guidance is opening new avenues to engage patients and healthcare providers (\cite{Alowais2023}, \cite{Secinaro2021}). Many healthcare systems in the developed world have been adopting these systems to increase patient satisfaction. One key shortcomings in this is that the majority of the developments in this domain are focused towards patients of European descent, their medical vocabularies. Many non-European languages, though spoken by millions, have seen very limited advancements. Bengali, despite being the seventh most popular language with 270 million speakers worldwide, has seen very limited progress in Bengali NLP and ASR research. This has hindered the integration of these technologies into digital health services for Bengali speakers which in turn slowed down the adoption of digital health solutions. While many European language speakers are benefiting from AI-driven services (conversational chatbot assisted) like digital appointment booking, symptom reporting before appointment and mental health support, Bengalis speakers are not able to benefit from these advancements.

Bengali ASR research has seen a significant surge in recent years, fueled by the release of large public speech corpora like Google's "Large Bengali ASR training data" (LB-ASRTD). This rich dataset has empowered researchers to explore advanced deep learning techniques like Baidu's Deep Speech 2 (DS2) and transfer learning approaches, yielding promising results with improved word error rates \cite{ShowravASR}. The focus extends beyond deep learning architectures, with studies investigating feature engineering advancements to enhance Bengali ASR performance. Hasan M. et al. explored the effectiveness of improved Mel-frequency cepstral coefficients (MFCCs) combined with deep Long Short-Term Memory (LSTM) networks for acoustic modeling. They further compare decoding strategies using a combination of Connectionist Temporal Classification (CTC) and a statistical language model (LM) with a CTC-based greedy decoder \cite{Hasan2019}. Additionally, Himadri Mukherjee et al. propose a Bangla phoneme recognition system using a novel feature extraction method called LPCC-2, achieving high accuracy on vowel phoneme recognition. This demonstrates the potential of improved feature representations for Bengali ASR tasks \cite{Mukherjee2018}.

Domain-specific automated speech recognition (ASR) models for medical applications in high-resource languages like English have shown significant promise in boosting healthcare delivery efficiency.  These models achieve greater accuracy compared to generic ASR models due to their tailored focus on medical terminology and vocabulary, and have demonstrably enhanced various healthcare functionalities \cite{healthASR1}, \cite{healthASR2}.

While recent advancements are in Bengali ASR are promising, significant hurdles remain before it can be effectively deployed in critical real-world domains such as healthcare. One major challenge is the  presence of numerous dialects and the influence of Sanskrit vocabulary create ambiguity for ASR systems, as words may be pronounced differently or share similar pronunciations with distinct meanings. This can lead to errors in interpreting spoken language, particularly in a sensitive domains such as healthcare, where misinterpretations can have serious consequences. This is further compounded by lack of domain specific quality labeled data. Unlike English and many European languages where vast amounts of conversation data for exist from sources like health insurance and healthcare providers, Bengali currently lacks such a comprehensive resource. This data scarcity hinders the development of robust ASR models that can accurately recognize medical terminology and nuances of healthcare conversations.

Medical domain-specific ASR systems hold immense potential for expanding digital health systems in LMIC countries like Bangladesh, which suffers from a low patient-doctor ratio.

\par \textbf{Voice-interactive remote patient triage:} AI-driven remote patient triage via digital health applications has gained significant popularity in recent years.  These applications collect patients' symptoms before appointments to reduce consultation time \cite{adahealth}  A Bengali medical ASR system can significantly improve accessibility for people with low digital literacy, removing a current barrier to these advancements.

\par \textbf{Transcription of doctor-patient consultations:} With a shortage of trained doctors, Bangladesh has one of the shortest average medical consultation durations globally \cite{Irving2017}. ASR systems offer a solution by transcribing and summarizing doctor-patient conversations, automatically increasing efficiency.

\par \textbf{Development of voice-based diagnostic tools:} ASR has the potential to play a crucial role in developing tools that analyze speech patterns for early detection of certain health conditions.

\par\textbf{Accessibility for Low Digital Literacy Users:} The healthcare sector shouldn't be limited to tech-savvy individuals.  ASR can empower people with disabilities or limited mobility by enabling interaction with medical equipment through voice commands.

This paper presents a proof-of-concept automated speech recognition (ASR) system specifically designed for Bengali biomedical data. We evaluate the performance of two popular ASR frameworks: DeepSpeech2 \cite{deepspeech2} and a fine-tuned Whisper BanglaASR. Both models are trained on a comprehensive Bengali medical corpus encompassing disease names, symptoms, and symptom severity. This corpus comprises 57.59 hours of data. Notably, both models achieve significant improvements in word error rate (WER) when transcribing medical conversations, with WER of 17.25\% and 9.05\% on a 5.8 hours of held-out test dataset. By addressing these challenges, we aim to bridge the gap in AI accessibility for the Bengali language and pave the way for the development of inclusive AI-powered healthcare solutions.

Model availability: The fine-tuned Whisper Bangla medical ASR model is currently deployed in AmarDoctor platform. Details of model testing data is available at: \url{http://amardoctor.health/banglamedasr.html}

\section{Methodology}

\subsection{Data Collection}

\begin{figure}[h]
    \centering
    \includegraphics[width=0.8\textwidth]{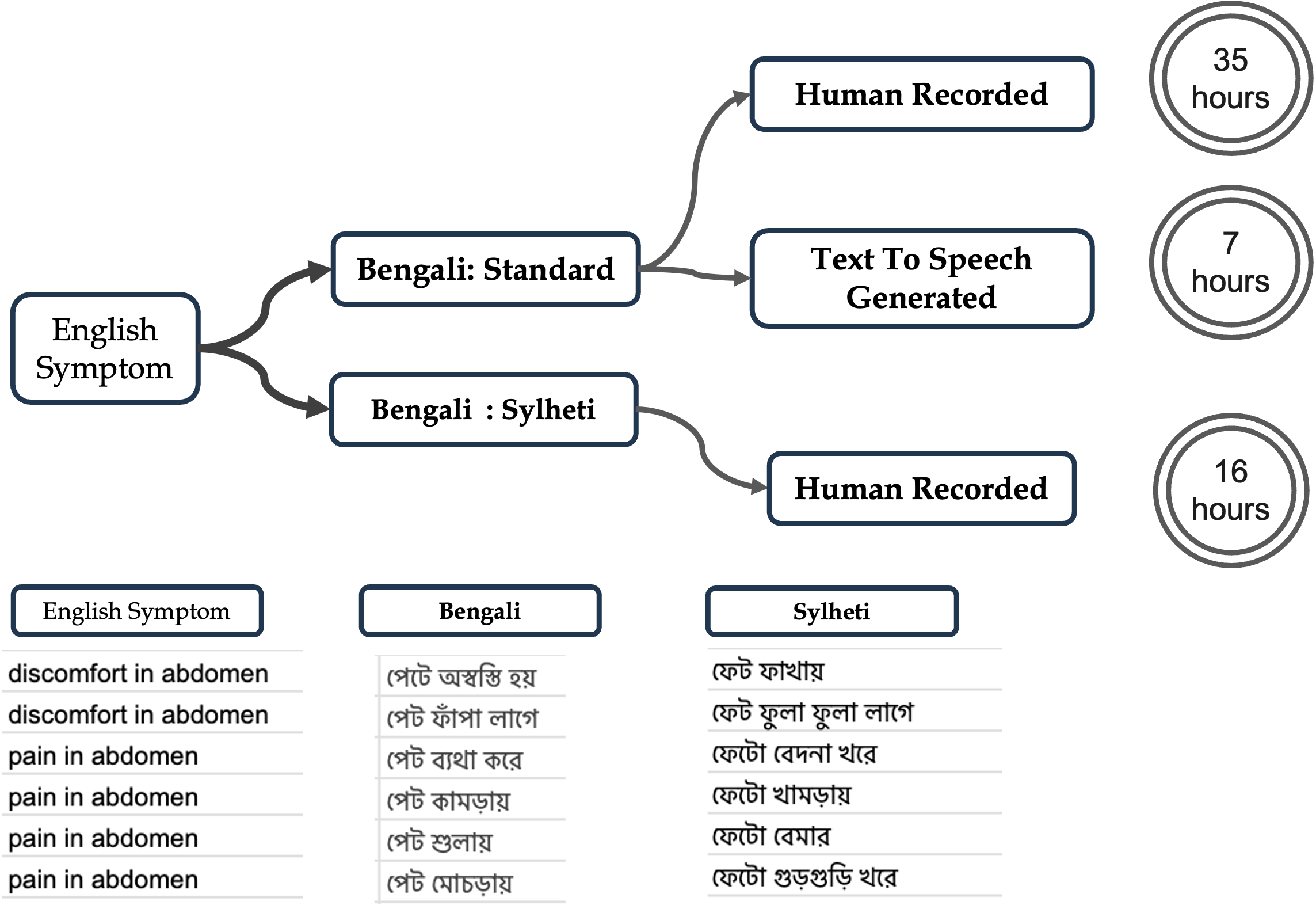}
    \caption{Distribution of the collected audio datasets.}
    \label{fig:data_collection}
\end{figure}

Our data collection process involved gathering various types of audio data related to medical symptoms, specifically designed to create a diverse and comprehensive dataset for Bengali medical ASR. This section provides a detailed overview of the data collection methodology:

We collected data from three distinct sources:

\textbf{Mapped Medical Symptoms:} We mapped 1,264 unique English medical symptoms to Bengali and Sylheti dialects \ref{fig:data_collection}. These symptoms were then incorporated into short, colloquial sentences. Audio recordings of human participants uttering these sentences were collected. Participants were recruited from a diverse range of demographics, encompassing gender (male and female) and age groups (young, middle-aged, and older adults).

\textbf{Synthetic Speech Data:} In addition to human recordings, we also collected synthetic data for the same sentences using Google Text-to-Speech. This approach helps to expand the dataset and introduces additional variation in pronunciation.

\textbf{Simulated Medical Conversations:} We further enriched the dataset by creating artificial medical scenarios. With informed consent, de-identified audio data was collected from real patient-doctor conversations based on these scenarios. This approach helps the model better understand the nuances of natural medical speech.

By combining these three data sources, we aimed to create a robust and versatile dataset that can effectively train our Bengali medical ASR system. The overall composition of our dataset is illustrated in Figure \ref{fig:data_collection}. This figure highlights the distribution of the data across different categories and languages, showcasing the diverse nature of the collected audio recordings.

\subsubsection{Data Sources}

We collected audio data from three primary sources:

\begin{itemize}
    \item \textbf{Human Recorded Symptom Data:} This dataset consists of recordings of medical symptoms spoken by native Bengali speakers. The recordings were further divided into two subsets based on the accent:
    \begin{itemize}
        \item \textbf{Standard Bengali:} Comprising 34.73 hours of audio data across 36,982 files.
        \item \textbf{Sylheti Bengali:} Consisting of 15.72 hours of audio data across 20,076 files, capturing the regional accent variations.
    \end{itemize}
    \item \textbf{Synthetic Data:} This dataset was generated using the Google Text-to-Speech API to synthesize medical symptom audio recordings. It includes 6.45 hours of audio data across 8,982 files.
    \item \textbf{Doctor-Patient Conversation Data:} We also collected 177 audio recordings of actual doctor-patient conversations, amounting to 0.68 hours. These recordings provide real-world interaction scenarios and enrich the dataset with natural dialogue patterns.
\end{itemize}

\subsubsection{Data Collection Summary}

Table \ref{tab:data_summary} summarizes the collected datasets, including the number of files and the total duration of each dataset.

\begin{table}[h]
    \centering
    \begin{tabular}{|l|c|c|}
        \hline
        \textbf{Dataset} & \textbf{Number of Files} & \textbf{Duration (hours)} \\
        \hline
        Standard Bengali & 36,982 & 34.73 \\
        Sylheti Bengali & 20,076 & 15.72 \\
        Synthetic Data & 8,982 & 6.45 \\
        Doctor-Patient Conversations & 177 & 0.68 \\
        \hline
        Total & 66217 & 57.59 \\
        \hline
    \end{tabular}
    \caption{Summary of collected datasets}
    \label{tab:data_summary}
\end{table}

\subsubsection{Data Statistics}

Table \ref{tab:data_stats} shows an analysis of the statistics of the audio files within our datasets:
\begin{table}[]
    \centering
    \begin{tabular}{|l|r|}
    \toprule
        Mean audio length &      3.130898 seconds \\
        Standard deviation of audio length   &      1.725558 seconds \\
        Shortest audio length   &      0.600000 seconds \\
        25\%   &      2.160000 seconds \\
        50\%   &      2.820000 seconds \\
        75\%   &      3.744000 seconds \\
        Longest audio length   &    150.048438 seconds \\
        
    \bottomrule
    \end{tabular}
    
    \caption{Basic statistics of collected data}
    \label{tab:data_stats}
\end{table}
This comprehensive data collection effort ensures that our dataset is well-suited for training and evaluating automated speech recognition systems in Bengali, with robust representation across different accents, recording methods, and interaction types.


\subsection{Data Preprocessing and Training}

Our data preprocessing and training pipeline involved several critical steps to ensure the quality and effectiveness of the automatic speech recognition (ASR) models. This section details the preprocessing methods and training procedures for both the DeepSpeech and Whisper models.

\subsubsection{Data Preprocessing}
\begin{figure}[!h]
    \centering
    \includegraphics[width=0.8\textwidth]{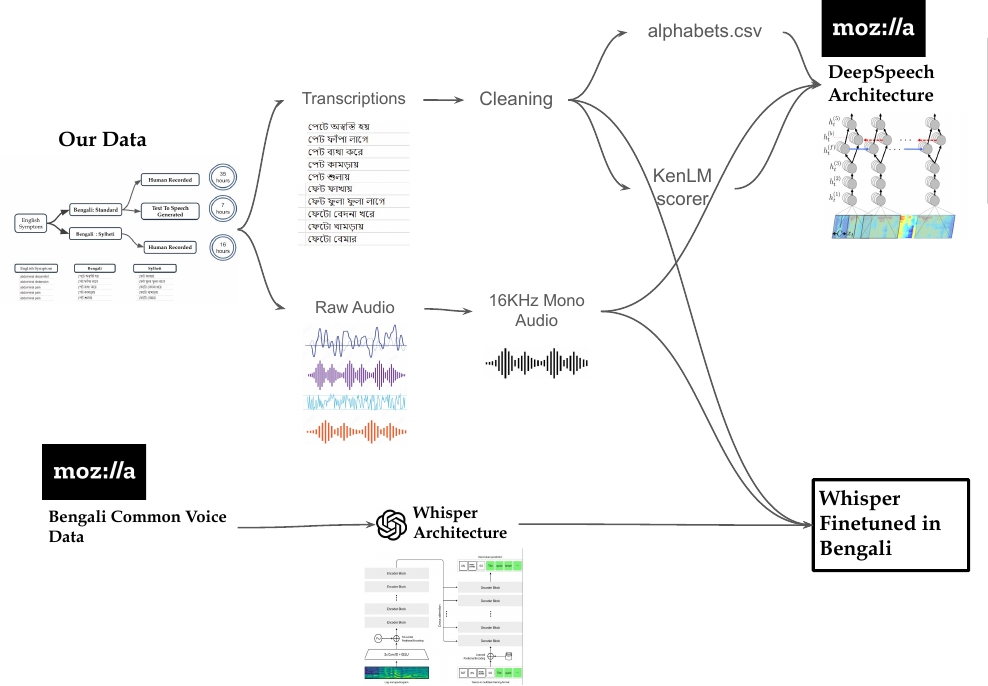}
    \caption{Data preprocessing and training pipeline for DeepSpeech and Whisper models.}
    \label{fig:training_pipeline}
\end{figure}

Prior to model training (DeepSpeech) or fine-tuning (Whisper), the collected audio data underwent a multi-step processing pipeline. Figure \ref{fig:training_pipeline} depicts the multi-step preprocessing pipeline employed to process the different datasets and our training architecture.

\paragraph{Audio Conversion}

All audio files were converted to 16 kHz single-channel WAV format. This standardization ensures uniformity across the dataset, reduces computational load, and maintains the audio quality necessary for effective speech recognition. The conversion to mono channel simplifies the audio input for the models, focusing on a single source of sound.

\paragraph{Transcription Normalization}

Transcriptions were normalized by removing any punctuation marks. This step helps in creating clean and consistent text data for training, which is crucial for the ASR models to learn accurate mappings from audio to text. Each audio file contains a single symptom spoken by one speaker/patient, ensuring that the dataset remains focused and consistent.

\paragraph{Alphabet Configuration}

An `alphabets.csv` file was used to define the set of characters used by the DeepSpeech model. This file includes all possible characters present in the transcriptions, ensuring that the model can handle the full range of phonetic and linguistic variations in Bengali.

\subsubsection{Noise Augmentation}

To enhance the robustness of DeepSpeech, various noise augmentation techniques were employed. These are explained in Table \ref{tab:noise_aug}.

\begin{table}[h]
    \centering
    \begin{tabular}{c|c|c}
        \hline
        \textbf{Name} & \textbf{Description} & \textbf{Probability} \\
        \hline
        Overlay augmentation & Layers another audio source onto the samples  & 50\% \\
        Warp augmentation & Applies a non-linear image warp to the spectrogram  & 10\% \\
        Reverb augmentation & Adds simplified Schroeder reverberation to the samples  & 10\% \\
        Frequency mask augmentation \cite{park2019specaugment} & Sets random frequency intervals to zero  & 10\% \\
        Resample augmentation & Resamples the samples to another rate and back  & 10\% \\
        Time mask augmentation & Sets random time intervals to zero  & 10\% \\
        Codec augmentation & Compresses and decompresses using Opus codec  & 10\% \\
        Dropout augmentation & Randomly drops segments  & 10\% \\
        Volume augmentation & Levels samples to a target dBFS value  & 10\% \\
        Pitch augmentation & Changes pitch by scaling spectrogram on the frequency axis  & 10\% \\
        Tempo augmentation & Changes playback tempo by scaling spectrogram on the time axis  & 10\% \\
        \hline
    \end{tabular}
    \caption{Applied Noise Augmentations}
    \label{tab:noise_aug}
\end{table}
\subsubsection{Model Training}

\paragraph{DeepSpeech:}
DeepSpeech, developed by Baidu lab \cite{deepspeech2}, was employed for training the ASR model. The architecture includes a Deep Neural Network (DNN) that converts audio into a sequence of probabilities over characters, followed by an N-gram language model (KenLM) \cite{heafield2011kenlm} that converts these probabilities into a sequence of characters.

\begin{itemize}
    \item \textbf{Training Data:} Our dataset was divided into training and validation sets, with specific focus on Symptom Data, Sylhet Data, and Synthetic Data.
    \item \textbf{KenLM Scorer:} The language model was built using transcriptions of our symptom data to improve accuracy. We evaluated the model's performance with and without the KenLM scorer.
\end{itemize}

\paragraph{Whisper:}
Whisper \cite{radford2023robust} is an advanced ASR model trained on 680,000 hours of multilingual and multitask supervised data collected from the web. We fine-tuned the Whisper model using the Mozilla Common Voice Bengali Dataset (MCVB) \cite{BanglaASR}, leveraging its robustness to accents, background noise, and technical language.

\begin{itemize}
    \item \textbf{Model Variant:} The Whisper small variant, with 244 million parameters, was selected for its balance between performance and computational efficiency.
    \item \textbf{Training Process:} The model was fine-tuned over 4K steps using 53K training samples and 6.6K validation samples, resulting in a word error rate (WER) of 3.22\% on the validation set.
\end{itemize}

\section{Experimental Results}

This section presents the Word Error Rate (WER) achieved by each model on various test datasets, including Symptom Data, Synthetic Data, Sylhet Data, Noisy Audio Data, and a Mixed Dataset (Combined Common Voice \& Symptom Data). WER is a common metric used in ASR tasks, reflecting the percentage of words incorrectly recognized by the model. Lower WER indicates better performance.

\textbf{Performance on Domain-Specific Data:}

Both DeepSpeech2 and fine-tuned Whisper BanglaASR significantly outperformed generic Bengali ASR models on the Symptom Data, as expected.  Wav2vec2 displayed the highest WER (most errors) in Sylheti Data (93.4\%), likely due to its dialectal limitations. Conversely, it achieved the lowest WER (fewest errors) in Synthetic Data (61.36\%), which may be attributed to the cleaner and controlled nature of synthetic speech.

DeepSpeech2 performed exceptionally well in Synthetic Data with a WER of only 2.07\%, demonstrating its proficiency in transcribing clear speech. However, it faced challenges with Sylheti Data, resulting in a higher WER of 22.30\%. This highlights the model's sensitivity to dialectal variations.

Whisper, when trained solely on generic Bengali data from Common Voice, struggled significantly with Sylheti Data, registering a WER of 110.01\%. This suggests that a generic model may not generalize well to specific domains or dialects without further adaptation. However, Whisper showed improved performance in Synthetic Data (46.46\%), indicating its potential with cleaner audio.

\textbf{Performance Across Different Data Types:}

When evaluating the models across the Mixed Dataset, Whisper achieved its best result in Synthetic Data (1.05\%), further confirming its strength with clean speech.  Conversely, its performance dropped considerably in Noisy Audio Data (20.21\%), suggesting a need for further noise reduction techniques.

These results highlight the importance of using domain-specific data for training ASR models. While DeepSpeech2 and fine-tuned Whisper achieved better overall performance with medical data, both models still face challenges with dialectal variations and noisy audio.

Table \ref{tab:wer_performance} presents the Word Error Rate (WER) for each model on the Symptom Data, Synthetic Data, and Sylhet Data, as well as the overall WER.

\begin{table}[h]
    \centering
    \begin{tabular}{|l|p{0.20\textwidth}|p{0.20\textwidth}|p{0.20\textwidth}|p{0.20\textwidth}|}
        \hline
        \textbf{Dataset} & \textbf{wav2vec2 on MCVB data} & \textbf{DeepSpeech2 on MCVB data \& Symptom} & \textbf{Whisper on MCVB data} & \textbf{Whisper on MCVB \& Symptom Data}\\
        \hline
        Symptom Data &71.35\% & 18.20\% & 58.74\% & 7.83\% \\
        Synthetic Data & 61.36\% & 2.07\% & 46.46\% & 1.05\% \\
        Sylhet Data & 93.45\% & 22.30\% & 110.01\% & 14.89\% \\
        Overall & 76.71\% & 17.25\% & 72.68\% & 9.05\% \\
        Noisy Audio & 89.91\% &41.7 \% & 73.69\% & 20.21 \% \\
        \hline
    \end{tabular}
    \caption{WER performance of DeepSpeech and Whisper models on different datasets}
    \label{tab:wer_performance}
\end{table}

\section{Discussion and Future Work}

The development of robust Automated Speech Recognition (ASR) systems for Bengali is crucial for improving accessibility and enabling AI-powered healthcare solutions in this region. Our work involves training and fine-tuning DeepSpeech and Whisper models on diverse Bengali speech datasets, including those with medical symptoms and different dialects. In this section, we discuss the performance of these models, particularly their handling of noisy audio, and outline future work to enhance their capabilities.

\subsection{Noisy Audio Issue}

One of the key hurdles hindering real-world deployment is the limited capability of both models to handle noisy audio data. While during DeepSpeech model training we have employed an extensive noise augmentation on training data, noise patterns available in these augmentation libraries were primarily reflecting noise patterns prevalent in standard developed countries, which often differ significantly from those encountered in South Asian environments. Consequently, neither model performed well on real conversation data from Bangladesh due to the mismatch between training noise and real-world noise profiles. Whisper, however, exhibited a slight advantage over DeepSpeech in noisy scenarios. This suggests potential for improvement by incorporating more real-world noisy data from Bangladesh during the fine-tuning process.

\subsection{Whisper Offers Superior Performance}

Among the evaluated models, Whisper fine-tuned on medical symptom data emerged as the clear leader. This superiority stems from its robust training on a vast multilingual and multi-task dataset, allowing it to handle diverse accents, noises, and speech patterns. Additionally, its training on a massive dataset (680,000 hours) encompassing various background noises contributes to its resilience in noisy environments. Finally, the significant performance improvement of the Whisper model fine-tuned on both MCVB (presumably Bengali speech data) and medical data compared to the generic data model underscores the critical role of domain-specific data for optimal ASR performance.

However, we still consider carrying forward the DeepSpeech model for several reasons:
\begin{itemize}
    \item \textbf{Customizability:} DeepSpeech is highly customizable. We have full access to its architecture, allowing us to tailor the model to the specific requirements of our application.
    \item \textbf{Licensing:} There are no licensing concerns with DeepSpeech, making it a viable option for commercialization and integration into proprietary systems.
    \item \textbf{Control Over Training:} Although DeepSpeech requires training from scratch, which demands more preprocessing and larger amounts of data, this process allows for fine-grained control over the training parameters and dataset composition.
\end{itemize}

Despite these advantages, it is clear that Whisper's pretraining on a massive dataset gives it a performance edge. As such, Whisper, being pretrained on a huge amount of data, will likely always outperform DeepSpeech in general scenarios.

\subsection{Key Limitations}
\begin{figure}[h]
    \centering
    \includegraphics[width=1\textwidth]{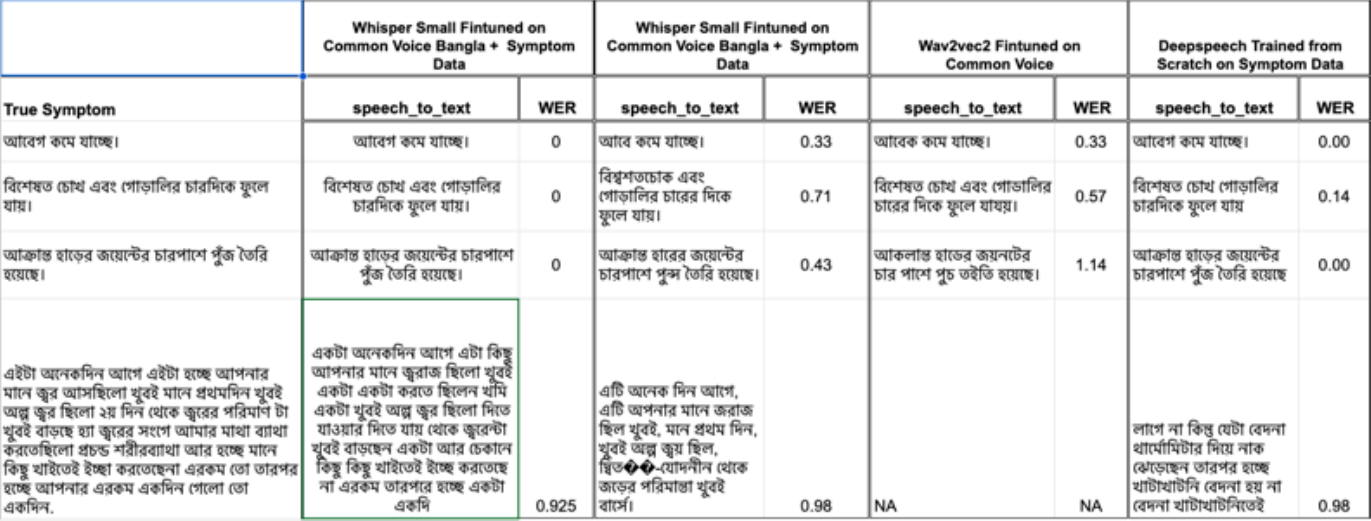}
    \caption{Example speech to text output of the compared models on held out test set.}
    \label{fig:example_output}
\end{figure}

This work emphasizes the critical role of domain-specific data in enhancing Bengali medical ASR performance. However, several limitations warrant further exploration. Firstly, noise resilience necessitates improvement, as real-world healthcare environments in developing countries frequently present acoustically challenging conditions. Secondly, the model exhibits limitations in handling extended conversational speech. Because current models are trained with short medical symptom descriptions, they are not adequately equipped to deal with longer conversations (Figure \ref{fig:example_output}, example 4). As a result, their application is limited in healthcare applications such as medical conversation chatbots. To bridge this gap, future endeavors should focus on incorporating robust noise reduction techniques and expanding the training data with extensive real-world conversational data collected from healthcare settings.

\subsection{Future Work}

Our future work aims to enhance the performance of both DeepSpeech and Whisper models, particularly in noisy environments and with diverse Bengali dialects:

\begin{itemize}
    \item \textbf{Improved Noise Modeling:} We will train DeepSpeech with better noise modeling techniques specifically designed for Bangladeshi environments. This involves collecting and profiling noise samples typical to Bangladeshi households and outdoor settings and incorporating these into the training process.
    \item \textbf{Fine-Tuning Whisper with Longer Conversation Data:} We plan to fine-tune Whisper with additional lengthy conversation data and noisy audio data from real-world Bangladeshi settings to further improve its robustness and accuracy in noisy conditions.
\end{itemize}

\section{Conclusion}
This work lays the foundation for a Bengali biomedical speech recognition system. By addressing dialectal variations and accessibility concerns, we aim to empower users with low digital literacy and pave the way for the development of inclusive AI-powered healthcare solutions in the Bengali language.



\bibliographystyle{unsrt}  
\bibliography{references}

\end{document}